\documentclass[12pt,a4paper]{article}

\usepackage{epsfig,graphicx}
                                                  
\usepackage{amsmath,amsfonts,amssymb}
\usepackage{wrapfig}

\newcommand{\unit}{\leavevmode\hbox{\small1\kern-3.6pt\normalsize1}}

\def\lsim{\raise0.3ex\hbox{$\;<$\kern-0.75em\raise-1.1ex\hbox{$\sim\;$}}}
\def\gsim{\raise0.3ex\hbox{$\;>$\kern-0.75em\raise-1.1ex\hbox{$\sim\;$}}}
\newcommand{\captions}{\sf\caption}

\def\bsg{$b\to s\gamma$}
\def\stau{\widetilde\tau_1}

\allowdisplaybreaks

\def\bsg{$b\to s\gamma$}
\def\stau{\widetilde\tau_1}

\def\estau{E_{\widetilde\tau}}
\def\esl{E_{\widetilde l}}
\def\esq{E_{\widetilde q}}
\def\mstau{m_{\widetilde\tau}}
\def\tstau{\tau_{\widetilde\tau}}

\begin{document}

\thispagestyle{empty}
\begin{flushright}
  FTUAM 08/21\\
  IFT-UAM/CSIC-08-76\\
  \vspace*{2.5mm}{5 December 2008}
\end{flushright}

\begin{center}


  {\Large \textbf{Stau detection at neutrino telescopes\\
                 in scenarios with supersymmetric dark matter} }  
  

  \vspace{0.5cm}
  Beatriz Ca\~nadas$^{1,2}$, David~G.~Cerde\~no$^1$,
  Carlos Mu\~noz$^1$, Sukanta Panda$^{1,3}$\\[0.2cm] 
    
  {$^{1}$\textit{Departamento de F\'{\i}sica Te\'{o}rica C-XI,
      and Instituto de F\'{\i}sica Te\'{o}rica UAM-CSIC,\\[0pt] 
      Universidad Aut\'{o}noma de Madrid, Cantoblanco, 28049 Madrid,
      Spain}}\\[0pt] 
  {$^{2}$\textit{INFN-Roma Tor Vergata, Via della Ricerca Scientifica
      1, I-00133 Roma, Italy}}\\[0pt]  
  {$^{3}$\textit{Indian Institute of Science Education and Research
      Bhopal, \\ 
      Govindpura, Bhopal 460 023, India}}\\[0pt]

  \vspace*{1cm}
  \begin{abstract}
    We have studied the detection of long-lived staus at
    the IceCube neutrino telescope,
    after their production inside the Earth through
    the inelastic scattering of high energy neutrinos.
    The theoretical predictions for the stau flux are calculated 
    in two
    scenarios in which the presence of long-lived staus is naturally
    associated to viable supersymmetric dark matter. Namely, we
    consider the cases with superWIMP (gravitino or axino) 
    and neutralino
    dark matter (along the coannihilation region).
    In both scenarios the maximum value of the stau flux turns out to
    be about 1 event/yr in
    regions with a light stau.
    This is consistent with light gravitinos, with masses 
    constrained by an upper limit which ranges from 0.2 to 15~GeV,
    depending on the stau mass. Likewise, it is compatible with
    axinos with a mass of about 1 GeV and a very low reheating
    temperature of order 100 GeV. 
    In the case of the neutralino dark matter this
    favours regions with a low value of
    $\tan\beta$, for which the neutralino-stau coannihilation region
    occurs for smaller values of the stau mass.
    Finally, we study the case of a general supergravity theory
    and show how for specific choices of non-universal soft parameters
    the predicted stau flux can increase moderately. 
  \end{abstract}
\end{center}

	
	\newpage

\section{Introduction}
\label{intro}

The existence of high energy neutrinos has been proposed in several
theoretical models. They are produced in astrophysical sources via 
collision of hadrons or of hadrons with the surrounding photons. 
Since neutrinos are not deflected by magnetic fields and do not lose
energy through interactions with the background while travelling from the 
sources to us, they carry all the relevant information about the 
nature of the astrophysical sources.
Detection of such neutrinos has been attempted in current 
experiments like AMANDA and will also be pursued in future kilometer
scale detectors such as IceCube at the south pole and KM3NeT at the
Mediterranean Sea.

Through inelastic scattering with nucleons ($N$) inside the Earth,
these high energy neutrinos  
can produce exotic particles which, if
charged and long-lived, may be detected in the above mentioned 
$\mathrm{km}^3$ Cerenkov detectors, thereby opening a window to new
physics. 
This idea was first proposed in \cite{Albuquerque2003} and further
studied in \cite{Ahlers2006,Albuquerque2006_1,Ahlers2006_rocks} within the context of a supersymmetric theory. 
In particular, they considered the process $\nu + N \rightarrow 
\widetilde{q}+\widetilde{l}$, assuming a spectrum where the lighter stau, 
$\stau$, is a long-lived NLSP to which the squark and slepton promptly 
decay. 
In addition, the possibility of producing and detecting these 
long-lived particles within the Universal Extra Dimension model 
has been recently explored \cite{Albuquerque2008}. 
In \cite{Albuquerque2003,Albuquerque2006_1} the number of staus that
might be detected in  IceCube was estimated by a Monte-Carlo
simulation for fixed slepton, stau and wino masses and for three
different squark masses.  
In \cite{Ahlers2006,Ahlers2006_rocks} an analytic 
estimation of this number was performed
for the SPS7 supersymmetry benchmark point and for a toy 
model with sparticle masses right above the experimental limits. 
An interesting related possibility, namely the production of long-lived
staus after the collision of high energy cosmic rays with nuclei in
the upper atmosphere, was studied in \cite{Ahlers:2007js} showing that 
staus arriving at large zenith angles could be detected in IceCube. 
Finally, stau detection taking into account other neutrino sources
which act as background for neutrino telescopes has also been explored
\cite{beacom}.

In this work we have further explored the possibility of producing 
staus inside the Earth and detecting them at neutrino telescopes. 
More specifically, we have probed the parameter space of the Minimal
Supersymmetric Standard Model (MSSM) from the point of view of a
supergravity theory where parameters are defined at the GUT scale.
Using the renormalization group equations to calculate the resulting
supersymmetric spectrum, we calculate for each point
the theoretical predictions for the stau flux at IceCube.
Furthermore, we have also investigated the implications
which arise when a supersymmetric solution to the problem of dark
matter is also imposed. Indeed, the lightest supersymmetric particle
(LSP), when neutral, is an excellent dark matter candidate
in R-parity conserving models.
We have contemplated two scenarios which provide an LSP dark matter 
candidate as well as a long lived stau when it is the next-to-lightest
supersymmetric particle (NLSP).
In the first of them, the LSP is the lightest neutralino 
\cite{neutralino}. In order to 
have a long lived enough NLSP in this case, a tight degeneracy between 
its mass and that of the stau
is necessary.
The second scenario is that with a superWIMP LSP, namely the 
gravitino \cite{gravitino} or the axino \cite{axino}. 
Both particles are characterized 
by extremely weak interactions, which entails a long lived
NLSP. 
In exploring these two scenarios, we have taken into account 
the most recent experimental constraints (such as limits on the 
masses of supersymmetric particles, and on low energy observables),
together with the present bounds on the relic density of cold dark
matter.

This paper is organized as follows. In Section 2 we describe the 
physical processes leading to slepton detection at neutrino 
telescopes. In Section 3, we describe in detail the two above mentioned
 scenarios, namely that where the neutralino is the LSP and that
with a superWIMP LSP. 
We then show the resulting theoretical predictions for 
stau detection rate in both of them. We also explore the
possible enhancement of the stau flux in scenarios with non-universal
soft supersymmetry-breaking parameters. Our conclusions are 
left for Section 4.

\section{Slepton detection at neutrino telescopes}

The existence of ultra high energy cosmic rays as well as the
detection of TeV photon emissions from galactic and extragalactic
sources (supernovae remnants, active galaxy nuclei and gamma ray 
bursts) are a strong 
indication for neutrino emission from the same sources.  
Waxman and Bahcall (WB) estimated an upper
bound for this flux assuming a proton cosmic ray flux proportional to
$E^{-2}$, motivated by first order Fermi acceleration models
\cite{wblimit1}. 
Mannheim, Protheroe and Rachen (MPR) also determined an upper limit for
diffuse neutrino sources \cite{MPRlimit} in almost the same way as WB,
but instead of assuming a specific cosmic ray spectrum, they defined
their spectrum based on current data at each energy. 
The resulting neutrino flux is approximately one order of magnitude
above the WB prediction. Notice that the bounds on the neutrino flux
could soon become more constrained by IceCube data \cite{:2007td}.

High energy neutrinos reach
the Earth at a given point on its surface, $l_{\oplus}$. As they
propagate through the Earth towards the detector,   
and due to Standard Model interactions, their flux will be attenuated.
Denoting the flux at the
Earth's surface as $F_{\oplus}(E_{\nu})$, 
then the flux at a position $l$ reads
\begin{equation}
  F_l(E_{\nu})=F_{\oplus}(E_{\nu})\ 
  \exp\left\{{\int_{l_{\oplus}}^{l}{\rho(l',\theta)
      \frac{\sigma^{SM}(E_{\nu})}{m_p}\,dl'}}\right\}\ ,
\end{equation}
where $\rho(l,\theta)$ is the Earth's density at each point, $m_p$ is
the proton mass, and $\sigma^{SM}$ corresponds to the neutrino-nucleon
scattering cross-section calculated in the Standard Model \cite{smcross}.
At this point, the neutrino interacts with a
nucleon, resulting in the production of a pair of supersymmetric
particles, which will further decay to staus and cross the
remaining distance to the detector, placed at $l=0$.

To calculate the number of events 
at the detector
per unit time and unit area, we have to multiply the parton level
supersymmetry cross-section ($\sigma^{SUSY}$), the corresponding
parton distribution function (PDF), 
$f(x,Q^2)$, the density at each point, 
and the neutrino flux taking into account its 
attenuation.
Upon integration in $l$, $\cos\theta$, $x$, $Q^2$ and neutrino energy,
the resulting flux (usually expressed in ${\rm yr}^{-1}{\rm
  km}^{-2}$) reads 
\begin{eqnarray}
  N=
  2\pi\int_0^1{d\cos\theta}\int_0^{l_{\oplus}} 
  {\frac{\rho(l,\theta)}{m_p}dl}\int_{x_{min}}^{1}{dx} 
  \int_{Q^2_{min}}^{Q^2_{max}}dQ^2 
 \int_{E_{\nu}^{min}}^{E_{\nu}^{max}} 
      {dE_{\nu}}f(x,Q^2)\frac{d\sigma^{SUSY}}
      {dxdQ^2}F_l(E_{\nu}) \ .
      \label{flux1}
\end{eqnarray}
The integration limits in $x$, $Q^2$ are given by kinematics, as well
as the lower limit of neutrino energy. 
For an upper limit we have chosen
$E_{\nu}^{max}=10^{10}$~GeV. Taking higher limits will not affect
the results, given the $E_{\nu}^{-2}$ dependence of the flux. The
limits on $l$ depend on three points: 
\begin{itemize}

\item Energy losses: 
  The average energy loss of a particle traversing a column depth $z$
  (where $dz=\rho(l,\theta)dl$) is given by
  \begin{equation}
    -\left\langle\frac{dE}{dz}\right\rangle=\alpha+\beta E\,,
    \label{energyloss}
  \end{equation} 
  where E is the energy of the particle, $\alpha$ describes ionization
  energy losses and $\beta$ is the radiative energy loss, which
  receives contributions from bremsstrahlung, pair production and
  photonuclear scattering.
  The parameter $\alpha$ is nearly constant as a function of the 
  mass of the particle, 
  $\alpha\approx2\times10^{-3}\ {\rm GeV\, cm^2/g}\,$.

  For low energies ($E\ll E^{cr}\equiv\alpha/\beta$), the range of 
  the stau is dominated by either ionization energy loss or by the 
  lifetime of the particle, and it scales linearly with energy.

  On the other hand, $\beta$ shows a dependence with both the energy 
  and the mass of the particle.  In particular, there is a 
  $1/m_{\widetilde{\tau}}$ dependence with stau mass, and an increase of 
  $\beta$ with stau energy. In this work we have used a 
  parametrization of $\beta$ obtained in \cite{energylosses1} by a 
  Monte-Carlo evaluation of the stau range including electromagnetic
  energy losses\footnote{
    The effect of weak interactions on the parametrization of
    $\beta$, especially those coming from charged current interactions
    can be comparable to that of electromagnetic interactions 
    \cite{energylosses2}. The impact on
    the range is maximal for pure left-handed mass eigenstates. It
    increases with the energy, and decreases with the mass of the
    stau. Given the $E^{-2}$ shape of the neutrino flux, detected
    staus are expected to come from neutrino interactions close to the
    threshold energy for squark production (i.e., approximately $10^6\,
    \mathrm{GeV}$). At that energy, these weak interaction effects are
    not important, and hence, in \cite{energylosses2} it is concluded
    that event rate estimates with energy losses parametrized as in
    \cite{energylosses1} are reasonably reliable. Moreover, the areas
    of the MSSM parameter space that we have explored give rise to
    a lighter stau with a small left-handed component.}.

  The dependence of $\beta$ with the mass of the particle is also true
  for muons, and the relation
  $\beta_{\mu}m_{\mu}\approx\beta_{\widetilde{\tau}}m_{\widetilde{\tau}}$
  holds. This means that $\beta_{\mu}$ is approximately three orders of
  magnitude  
  bigger than $\beta_{\widetilde{\tau}}$, causing the muon range to be 
  much smaller than the stau range. This crucial fact, 
  allows stau detectability even though their production 
  cross-section is about three orders of magnitude smaller than 
  standard model processes 
  \cite{Albuquerque2003}.

  \item Stau lifetime: If the stau is not long-lived enough, it may 
  decay before it reaches the detector. We will further comment on 
  this issue when we discuss in more detail the two scenarios we have 
  studied.

  \item Track separation: The angle between the two staus has to be 
  wide enough to ensure the detector to be able to discriminate 
  the arrival of the two particles from a one-particle event. We 
  therefore require that the particles are at least 50~metres apart at
  arrival. 
  On the other hand, if the angle is too wide, one of them, or both, 
  may miss the detector. Hence, we demand them to be at most 1 km
  apart from each other.

\end{itemize}

In principle, muons produced by upgoing atmospheric neutrinos could
mimic the signal of individual staus traversing the detector. This
kind of background is elliminated with the requirement that two
simultaneous tracks have to be observed in the detector.

This leaves the production and subsequent detection of a muon pair,
$\mu^+\mu^-$, as the main source of background. 
The stau track separation is a key
point in order to discriminate stau pair flux from this signal 
\cite{Albuquerque2006_1}. Due to their
shorter range, the detected muons are only those produced in the
vicinity of the detector. Hence, the separation of the tracks of
a $\mu^+\mu^-$ pair is
smaller or of the order of $100$~metres. On the contrary, staus can be 
produced at much larger distances and as a consequence their track
separation can be as large as several hundreds or even thousand of
metres.  
Thus, 
although the dimuon flux largely exceed the stau flux, for track
separations above $100$~m there should not be any significant
contribution from this background. Moreover, making use of the energy
deposition of the events it is possible to further reduce the dimuon
background. All these issues are 
explained in detail in Ref.\,\cite{Albuquerque2006_1}.

Hence, the limits on the range $z(\cos\theta,l)$ are calculated as
follows.  The upper limit, $z_{max}$, corresponds to 
the minimum between the particle's range due to energy losses
and the distance it can travel before decaying, $l=\gamma c \tau$.
The lower limit is determined by the requirements on the track 
separation: for a given angle, $l$ has to be such that
$\Delta=l\tan\theta_{LAB}$ is between $50$~m and $1$~km.

The analysis of particle tracks in neutrino telescopes is calibrated
for muons. The energy of the incoming muons that reach the IceCube
detector can be reconstructed through their energy loss $\Delta E$ per
column depth $\Delta z$, provided that
$E_{\mu}>E_{\mu}^{cr}$. According to equation (\ref{energyloss}), 
$\Delta E/\Delta z = \alpha + \beta_{\mu}E_{\mu}$. Thus staus
would be detected as muons with reduced energy $E_d\equiv
E_{\tilde{\tau}}\,m_{\mu}/m_{\tilde{\tau}}$, since
$\beta_{\tilde{\tau}}=\beta_{\mu}\, m_{\mu}/m_{\tilde{\tau}}$. We will use
this reduced energy to calculate the total rate of events 
at IceCube, as the acceptance of the telescope highly depends
on detected energy, as well as on the incoming direction of the
particles. Note that 
the critical energy for staus is much higher than that for muons and 
so for arrival energies below $E_{\widetilde{\tau}}^{cr}\sim
10^5$~GeV, one has $\Delta E/\Delta z\approx \alpha$ and hence it
would not be  possible to estimate their energies.

Finally, in our calculation we have used the effective area for upward
going muons, determined by the IceCube collaboration \cite{aeff},
averaged in the angular direction as it is detailed in Table 1 of 
Ref.\,\cite{Ahlers2006}. 
This implies a suppression factor in Eq.(\ref{flux1}), which is 
smaller than 1.25 for detected energies above $1000$~GeV and which
can be as large as a factor $10$ for energies below $100$~GeV. 
We have computed the effect of weighting with the IceCube effective 
area for a representative set points in the parameter space 
and have found it to induce a reduction in the total flux of
approximately a factor of $1.5$. This can be qualitatively understood
from the fact that 
the detected stau energy distribution peaks at approximately
$1000$~GeV \cite{Ahlers2006}.

\section{Results}

We are now ready to determine the theoretical predictions for the flux
of staus that could be observed in neutrino telescopes. In doing so,
we will work within the context of a supergravity theory, in which the
set of soft supersymmetry-breaking parameters are considered as inputs
at a high energy scale, which we will take as the scale at which gauge
couplings unify, $M_{GUT}\approx2\times10^{16}$~GeV. The
renormalization group equations (RGEs) for the MSSM are then
numerically solved to evaluate these parameters at the electroweak 
scale where
the supersymmetric spectrum is calculated. 
The minimization of the Higgs potential leaves the
following condition to be satisfied by the Higgsino mass parameter at
the SUSY scale, 
\begin{eqnarray}
  \mu^2 & = & 
  \frac{ - m_{H_u}^2\tan^2\beta + m_{H_d}^2}{\tan^2\beta-1} 
  - \frac{1}{2} M_Z^2 ,
  \label{muterm}
\end{eqnarray}
in terms of the ratio of the vacuum expectation values of the Higgs
doublet, 
$\tan\beta\equiv\langle H_u^0\rangle/\langle H_d^0\rangle$, which is
also considered an input in our analysis. Notice that the sign of
$\mu$ is also left undetermined.

Inelastic scattering on nucleons is the dominant interaction process
of high energy cosmic neutrinos both in the atmosphere and the inside
of the
Earth. The products of this interaction are sleptons and squarks which
can ultimately decay into the lighter stau, when this is the NLSP and
long-lived enough. The leading processes in the MSSM involve exchange of
charginos and neutralinos along a $t$-channel. 
The corresponding expressions of the 
differential cross-section for neutrino-quark
scattering into a pair of sleptons and squarks are explicitly shown in
the Appendix.

The theoretical predictions for the resulting stau flux are obviously
dependent on the initial
structure of the soft terms, which is a function of the (yet unknown)
mechanism of 
supersymmetry breaking. In the following study we will start by
assuming universality of the soft-parameters at the GUT scale,
studying the so called Constrained MSSM (CMSSM). The CMSSM is
fully specified by a common scalar
mass, $m$, a common gaugino mass, $M$, a trilinear parameter, $A$, the 
sign of the $\mu$ term and $\tan\beta$.

In exploring the supersymmetric parameter space we will impose several
experimental constraint in order to guarantee phenomenological
consistency. More specifically, 
we will consider the LEP bounds on the masses of
supersymmetric particles, as well as on the lightest Higgs boson. 
Moreover, we will also include the current experimental bound
on the branching ratio of the $b\to s\gamma$ decay, which sets the
most stringent constraints in the scenarios we analyse.
In particular, we will impose
$2.85\times10^{-4}\le\,{\rm BR}(b\to s\gamma)\le 4.25\times10^{-4}$,
obtained   
from the experimental world average 
reported by the Heavy Flavour Averaging Group \cite{bsgHFAG07},
and the theoretical calculation in the Standard Model
\cite{bsg-misiak}, 
with errors combined in quadrature.

As already explained in the introduction, we will consider two main
scenarios in which the presence of long-lived staus is well
motivated and associated to solutions to the problem of
the dark matter. In particular, we will start by analysing the CMSSM
scenario with a supersymmetric superWIMP (gravitino or axino LSP) 
and then we will extend our study to the
case of neutralino dark matter in the coannihilation region with the
stau. 
Finally, we will explore the effect of non-universalities in the soft
supersymmetry-breaking parameters on the predictions for the stau
flux.

\subsection{superWIMPs}

Two possible superWIMPs \cite{superwimps,axino2}
are viable dark matter candidates within the
framework of supersymmetric theories, namely the gravitino 
and the axino. 
In both cases the stau, when it is the NLSP, is long-lived
(easily exceeding $\tstau=10^{-9}$ s) due to the smallness of the
couplings (gravitational and Peccei-Quinn scale, respectively) governing
its decay into the LSP. Therefore, both situations are suitable frameworks
that would give rise to a flux of staus in neutrino Cerenkov
detectors. 
Let us
first briefly comment on both possibilities.

On the one hand, 
the gravitino, when it is the LSP in a supergravity scenario, can be
an excellent candidate for dark matter
 \cite{superwimps}.
Gravitinos can be thermally produced 
during the reheating of the
Universe, by ordinary processes involving scattering and decays of
particles in the primordial plasma. 
Besides, a
non-thermal production class of process 
also exist
when the NLSP has a
lifetime such that, being shorter than the age of the Universe, it is
still long enough so that it decouples from the plasma before
decaying to the gravitino LSP.  
Which contribution to the relic density is more significant depends on
the reheating temperature and on the gravitino mass, but in general
both have to be considered \cite{fujii,rosz05}.

In late decays of the NLSP into the LSP, 
additional electromagnetic and hadronic
showers are produced. If the decay takes place after Big Bang
Nucleosynthesis (BBN), the products of these showers may alter the
abundances of light elements \cite{gravitinobbn}.  
Moreover, late injection of electromagnetic energy may distort the 
frequency dependence of the cosmic microwave background 
spectrum from its observed blackbody 
shape \cite{hu,gravitinoTP1,pdg02}. 
Preventing these effects leads to constraints on the supersymmetric 
parameter
space that have to be imposed in addition to the usual experimental
bounds.
Recently, it has been pointed out \cite{CBBN1} that metastable charged
particles 
can form bound states with light nuclei, opening new
channels for thermal reactions and enabling the catalyzed BBN. 
In this case $^6$Li and $^9$Be overproduction becomes an issue and
places very strong upper bounds on staus lifetimes, 
$\tstau\lsim 5\times10^{3}$~s (see, e.g.,
\cite{cbbn2,Pospelov:2008ta}).

These constraints exclude extensive areas of the parameter space. 
In particular, in the case of the CMSSM and for moderate gravitino
masses, they disfavour 
the regions with
neutralino NLSP \cite{gravitinoCMSSM,gravitinoD,gravitinocmssm3}, 
leaving only some regions in which the stau is the
NLSP. Therefore, the presence of long-lived staus in the case of
gravitino LSP is very natural. 
Indeed, in this scenario, the stau decays to the gravitino and a
$\tau$ lepton at tree level, via gravitational interactions 
with a lifetime \cite{lifetau,gravitinoCMSSM}, 
\begin{equation}
  \tau_{\widetilde{\tau}}\approx
  6.1\times10^{4}\left(\frac{m_{\widetilde{G}}}
  {1\,\mathrm{GeV}}\right)^2 
  \left(\frac{100\,\mathrm{GeV}}{m_{\widetilde{\tau}}}\right)^5
  \left(1-\frac{m_{\widetilde{G}}^2}
       {m_{\widetilde{\tau}}^2}\right)^{-4}\mathrm{s}\ .
       \label{tautausecs}
\end{equation}

On the other hand, the axino (fermionic supersymmetric partner of
the axion) is another possible dark matter candidate \cite{axino2}. 
Due to 
the smallness of its coupling to ordinary matter, $1/f_a$ (with
$f_a\sim 10^{11}$\,GeV being the Peccei-Quinn scale), it has similar
properties as those of the gravitino. For example, the axino relic
density also 
receives contributions both from thermal and non-thermal production
processes. An important difference is, however, that the lifetime of
the NLSP is considerably smaller (since the axion coupling is much
smaller than the gravitational one). Hence, the NLSP typically decays
before BBN and the parameter space is generally free from constraints
on the abundance of light elements.

We proceed now to analyse the detection properties of the stau in
neutrino telescopes when it is the NLSP in the case of either the
gravitino or axino dark matter scenario in the CMSSM framework. 
We will therefore analyse those points in the CMSSM where the stau is
the lightest {\em observable} supersymmetric 
particle (LOSP) assuming that the rest of
the parameters (gravitino or axino mass and reheating temperature) can
be chosen in order to achieve viable gravitino or axino dark
matter 
\footnote{This might not be possible for all the points in the
  parameter space, but given a specific scenario for gravitino or
  axino dark matter 
  the corresponding BBN constraints and regions with correct DM relic 
  density can easily be superimposed on our plots.}.

\begin{figure}[t]
  \begin{center}
    \epsfig{file=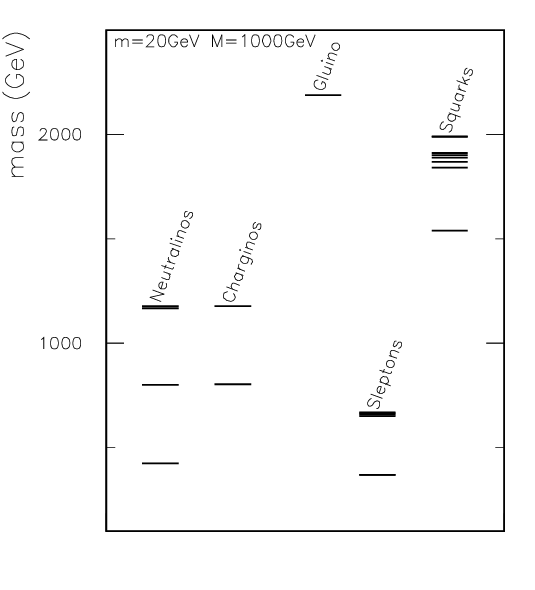,width=8.6cm}
  \end{center}
  \vspace*{-1cm}
  \captions{Supersymmetric spectrum corresponding to the CMSSM
    with
    $M=1000$~GeV, $m=20$~GeV, $A=0$ and 
    $\tan\beta=10$.  
  }
  \label{spectrum_univ}
\end{figure}

Thus, for each point in the parameter space we have calculated, using
the expressions in Appendix~\ref{appendix}, the sfermion production
from neutrino inelastic scattering inside the Earth. 
We have then used the code ISAJET \cite{isajet}
to check the subsequent decay
chains and branching ratios of these parent sleptons and squarks into
the lighter stau and we
have made an  
estimation of the average 
energy of the parent particle carried by the staus.
In particular, we have found that the following 
relations 
\begin{eqnarray}
  \langle\estau\rangle\approx\frac{1}{2}\,\langle\esl\rangle\ ;
  \quad \quad  
  \langle\estau\rangle\approx\frac{1}{3}\,\langle\esq\rangle\ ,
  \label{energy-universal}
\end{eqnarray}
hold for the entire
region of the CMSSM parameter space where the stau is the LOSP
with $M\lsim2$~TeV. This is similar to previous estimations
performed for the SPS7 benchmark point \cite{Ahlers2006}. For
completeness, we show in Fig.\,\ref{spectrum_univ} a representative
spectrum obtained in the CMSSM for $M=1000$~GeV, $m=20$~GeV, $A=0$ and 
$\tan\beta=10$.

Our choice for PDF's in Eq.(\ref{flux1}) corresponds to those obtained 
from the Cteq6PDF package \cite{cteq6}. However, in order to evaluate
the dependence of the resulting flux on the particular choice of 
PDF's, we have repeated our calculations using also the MRST2004
package \cite{Martin:2006qz}. We find a very small variation in the
predicted stau flux, that we can quantify as approximately a $2\%$.
Moreover, in the computation with the Cteq6PDF package, 
we have also evaluated the uncertainty in the resulting stau flux
which is due to the PDF's error. Following the procedure detailed in
Ref.\cite{cteq6} we have found that the error is approximately a
$2\%$.

Using then the procedure sketched in the previous Section, 
we have computed the flux of staus at the IceCube detector. We have
explored a slice of the CMSSM parameter space, setting $A=0$, $\mu>0$,
and varying the scalar and gaugino mass parameters, $m$ and $M$, in
the range $0-2$~TeV, retaining only those points where the stau is the
LOSP and the various experimental constraints are fulfilled.  

\begin{figure}
  \epsfig{file=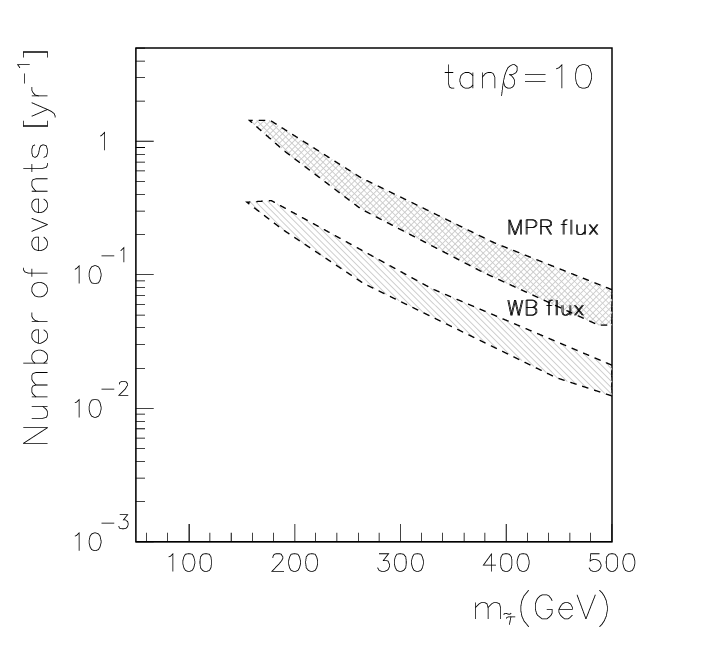,width=8.6cm}
  \hspace*{-1cm}
  \epsfig{file=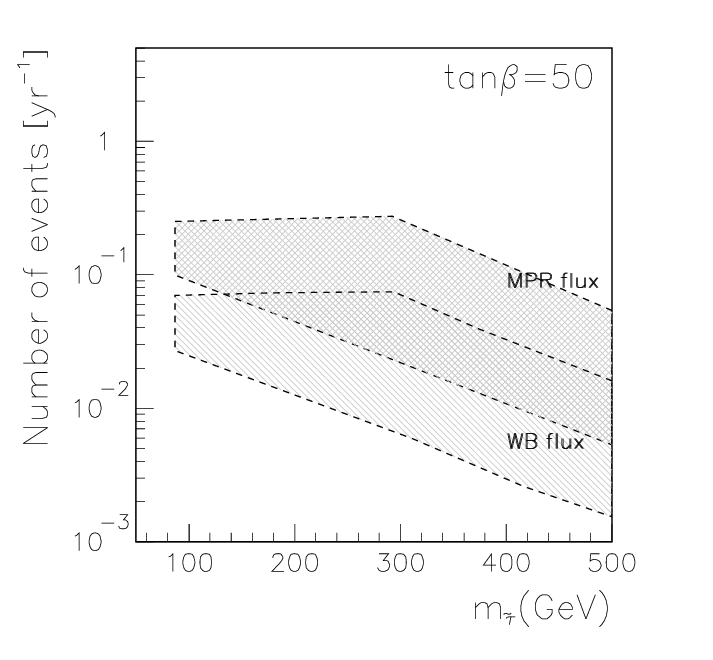,width=8.6cm}
  \vspace*{-1cm}
  \captions{Theoretical predictions for the stau flux at IceCube
    as a function of the stau mass for different values of
    $\tan\beta$ in the CMSSM scenario with $A=0$. } 
  \label{fluxewimp1}
\end{figure}

In Fig.~\ref{fluxewimp1}, we represent the stau flux as a function of 
stau mass for both WB and MPR neutrino fluxes and for 
$\tan\beta=10$ and $50$.
Let us first examine the case $\tan\beta=10$. As expected, the 
production of heavier staus is suppressed, and therefore the flux
decreases with the stau mass. We obtain, however, a band and not a line
when we plot flux versus mass. This can be explained in the 
following way. Neutrino interactions leading to stau pair production
involve the production of a squark, whose mass determines the energy
threshold of the whole process. Squark masses are strongly dominated
by gluino masses, which are roughly proportional to the common gaugino 
masses. 
Therefore, for the same value of the stau mass, the points on the
$(m,M)$ plane corresponding to lower values of $M$ give rise to a 
lighter spectrum and thus lower energy thresholds.
This,
together with the $E^{-2}$ shape of the neutrino flux, implies a 
higher number of stau pairs for lower values of $M$. Turning to the
figure, it is now easy to understand that, given a value for
$m_{\widetilde{\tau}}$,  
the highest value of the flux corresponds to the lowest 
value of $M$ for which the stau is the LOSP. 
The lowest value, in turn, corresponds to 
the highest value of $M$, i.e., that corresponding to $m=0$.
In the figure, the lower limit on the stau mass corresponds to the
LEP experimental constraint on the Higgs (stau) mass for
$\tan\beta=10$ (50).

For the $\tan\beta=50$ case, the flux is lower than that 
for $\tan\beta=10$ in approximately one order of magnitude. 
Notice that 
when $\tan\beta$ increases, the off-diagonal elements of the 
sparticle mass matrices become larger and the L-R mixing becomes 
more important. This implies a larger mass gap between the two mass
eigenstates of the third family. 
Thus, for a fixed stau mass, the rest of the spectrum is heavier
for $\tan\beta=50$ than for $\tan\beta=10$.
This, according to the previous
threshold energy argument, implies a lower flux.
Notice finally that for $\tan\beta=50$ the experimental constraint on
the branching ratio of the \bsg, which implies a lower bound on the
common gaugino mass, leads to an upper bound on the predicted stau
flux of about $0.2$ event yr$^{-1}$ for the MPR flux.

Thus the most optimistic results are
obtained for light staus, and the lowest possible stau mass
in the $\tan\beta=10$ scenario
sets an upper limit to the predicted flux of about 1
event yr$^{-1}$ when the MPR flux is used (with the WB prediction this
is reduced to $0.3$ event yr$^{-1}$).
Notice that these results are smaller than those obtained in
\cite{Albuquerque2006_1}
due to the more constrained framework of supergravity analyses. 
Despite the fact that, as explained in the previous Section, 
the stau events can be distinguished from the
dimuon background, these results evidence that, at best, 
several years of data from IceCube would be
necessary to claim a positive signal.
We will later see how this is modified when non-universal soft
parameter are included.

\begin{figure}[!t]
  \epsfig{file=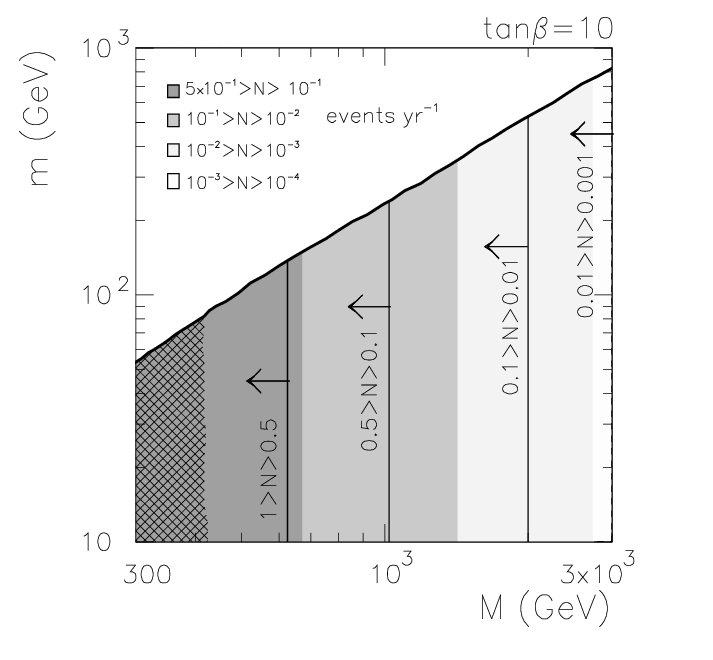,width=8.6cm}\hspace*{-1cm}
  \epsfig{file=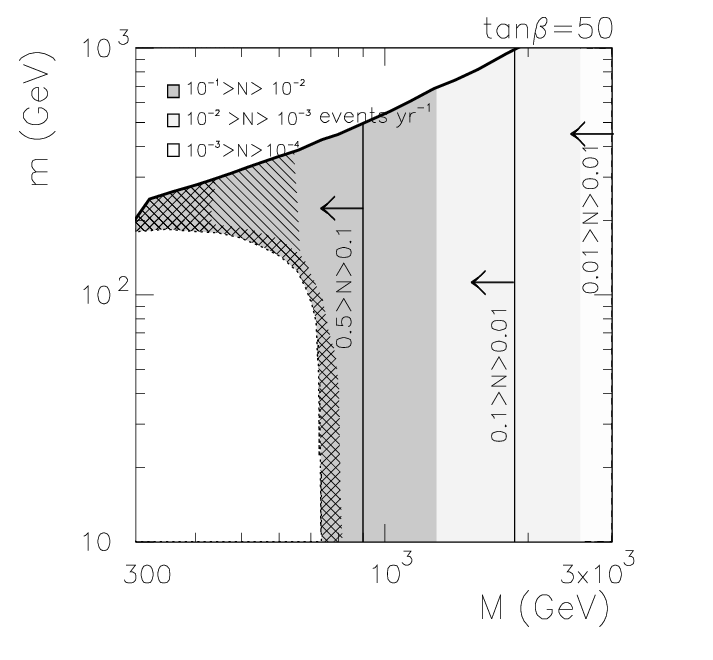,width=8.6cm}
  \vspace*{-1cm}
  \captions{$(m,M)$ plane of the CMSSM scenario 
    with $A=0$ and $\tan\beta=10,\,50$. The region below the thick
    solid line corresponds to that with 
    stau LOSP. Coloured regions represent the
    theoretical predictions for the stau flux at IceCube 
    in the case of WB neutrino flux, whereas vertical solid lines
    correspond to the results when the MPR flux is used. Gridded areas
    are experimentally excluded by LEP constraints on the Higgs and
    stau masses, whereas the ruled area is excluded by 
    $b\rightarrow s\gamma$. The white area on the lower left corner
    of the plot for $\tan\beta=50$ is excluded due to the occurrence
    of negative stau squared-mass.}
  \label{fluxewimpmM}
\end{figure}

In order to understand which regions of the CMSSM parameter space
could be explored using this technique, 
in Fig.~\ref{fluxewimpmM}
we superimpose the
theoretical predictions for the stau flux on the $(m,M)$ plane for
$A=0$ and $\tan\beta=10,\,50$. 
The most optimistic predictions regarding detectability occur for 
low values of $M$, corresponding to the regions with lighter
staus and the flux decreases as the gaugino mass increases, 
with almost no
dependence on the scalar mass parameter (when
the stau mass is RG-evolved, the main contributions to it come from
terms depending on gaugino masses). 
When the MPR prediction for the neutrino flux is used, a significant
increase of the resulting stau flux is obtained.

To sort out the limits on the stau lifetime imposed by catalyzed
BBN, the mass of the gravitino is constrained by an upper limit,
\begin{equation}
m_{\widetilde{G}}\leq 0.28\,\left(\frac{m_{\widetilde{\tau}}}{100\,
  \mathrm{GeV}}\right)^{5/2} \mathrm{GeV}\,, 
\end{equation} 
which can be deduced from Eq.(\ref{tautausecs}), 
setting the stau life
equal to $5\times10^3$ s. It can then be easily seen that the
maximum gravitino mass ranges approximately from $0.2$~GeV to 
$20$~GeV, depending on the mass of the stau. For such light 
gravitinos, the relic density is fully dominated by thermal 
production,
\begin{equation}
  \Omega^{TP}_{m_{\widetilde{G}}}\,h^2\approx0.27\,
  \left(\frac{T_R}{10^{10}\,\mathrm{GeV}}\right)
  \left(\frac{100\,\mathrm{GeV}}{m_{\widetilde{G}}}\right)
  \left(\frac{m_{\widetilde{g}}(\mu)}{1\,\mathrm{TeV}}\right)^2,
  \label{omegatp}
\end{equation} 
where $m_{\widetilde{g}}(\mu)$ is the running gluino mass 
\cite{Bolz:2000fu,Pradler:2006qh}. 
A relic density
of about $0.1$, in agreement with WMAP 
data \cite{wmap}, can be recovered by an appropriate choice of the reheating
temperature, depending on the mass of both the gravitino and the
gluino. As $T_R\propto m_{\widetilde{G}}/m_{\widetilde{g}}^2$, we can 
make a rough estimation of the maximum value of $T_R$ as a function 
of the stau mass,
\begin{equation}
  T_R\lesssim 5.2\times
  10^7\left(\frac{m_{\widetilde{\tau}}}{100\,\mathrm{GeV}}
  \right)^{1/2} \mathrm{GeV}. 
\end{equation}
Similar upper bounds for the reheating temperature 
have been derived \cite{gravitinocmssm3,cbbn2,tr} and used to
study its possible determination at the LHC \cite{tr}.
It is finally worth pointing out that 
in this case the contribution to $\Omega_{DM}$  
from non-thermal production is negligible.

As in the case of the gravitino, axino thermal production is a
function of the reheating temperature
\cite{axino2,Brandenburg:2004du}. 
As emphasized in
\cite{axinocmssm}, the correct relic density can be obtained for 
nearly any point in the $(M,\,m)$ by fixing the axino mass and the
reheating temperature. In particular, the regions with a small value
of the gaugino mass (where the stau flux is maximal) could correspond
to an axino with mass $m_{\widetilde a}\sim{{\cal O}(1\,{\rm GeV})}$
which must have a small 
reheating temperature, $T_R\sim200\,{\rm GeV}$, 
in order to satisfy the constraint on its relic abundance. 
Notice that although a gauge invariant treatment of the axino thermal
production \cite{Brandenburg:2004du} is only valid for larger $T_R$, 
this value can be obtained within the context of a quitessential
kination scenario \cite{Gomez:2008js}. 
BBN
constraints in this case
are easily fulfilled since the stau lifetime is usually smaller than
1 second.

\subsection{Neutralino LSP}

Let us now consider a second interesting class of MSSM scenarios,
those with a neutralino LSP as the dark matter candidate. 
Under these circumstances, 
in order to have a sufficiently long lifetime, the 
stau NLSP needs to be almost degenerate with the lightest neutralino.
Interestingly, the quasi-degeneracy of the stau and lightest neutralino
is also welcome from the point of view of neutralino dark
matter. In such a situation the relic abundance of neutralinos is
largely suppressed through a coannihilation mechanism \cite{coann},
which makes it possible to find agreement with the constraints on
the dark matter abundance \cite{coann2}.
Thus, this is another example in which supersymmetric
scenarios which solve the dark matter problem can naturally provide
long-lived staus.

The dependence of the stau lifetime 
on $\Delta m=m_{\widetilde{\tau}}-m_{\widetilde{\chi}^0}$ and on the left-right 
content of the mass eigenstate was studied in \cite{staulife}.
It was shown that
in the regions where $\Delta m>m_{\tau}$, 
the 2-body decay $\widetilde{\tau}\rightarrow\tau\widetilde{\chi}^0$ is
allowed, and it is the dominant process. 
In this case, the stau decays very 
rapidly, with a lifetime smaller or of the order of $10^{-17}$~s, and
therefore never reaches the detector. 
On the other hand, if $\Delta m < m_{\tau}$, this
channel closes, and typical lifetimes 
become $\mathcal{O}(10^{-6}\, \mathrm{s})$ or greater. This is long 
enough for staus to reach the detector before decaying
\footnote{For example, 
  given the mean free path for a stau, $l_{\widetilde\tau}
  =c\,\tstau\,\estau/\mstau$, imposing $l_{\widetilde\tau}\gsim 1$ km 
  for a stau of $\mstau\sim 150$~GeV and
  an energy of $10^6$~GeV, this would imply a lifetime of at least
  $10^{-9}$~s.}.

\begin{figure}[t]
  \epsfig{file=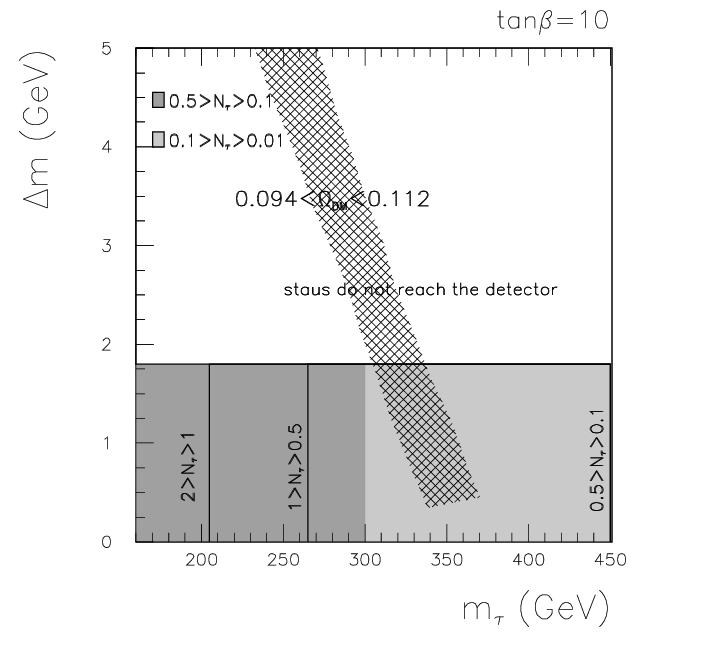,width=8.6cm}\hspace*{-1cm}
  \epsfig{file=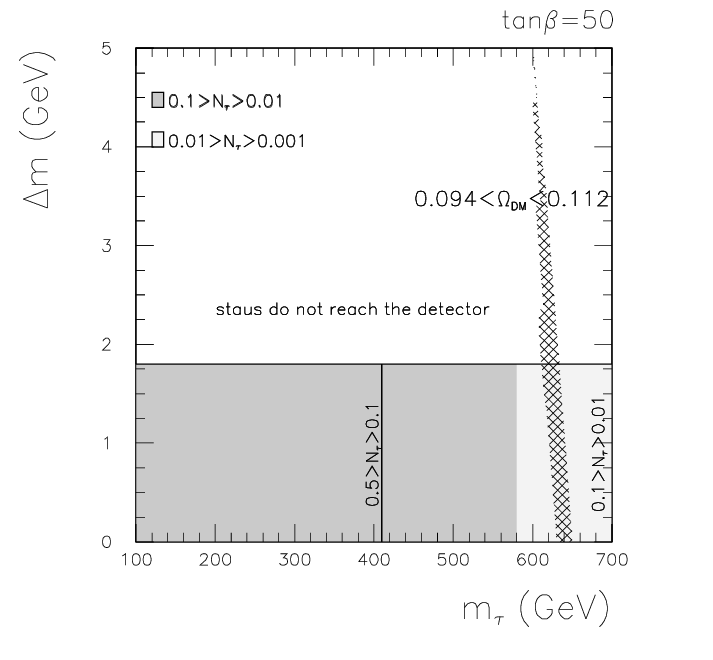,width=8.6cm}
  \vspace*{-1cm}
  \captions{$(\Delta m,m_{\widetilde{\tau}})$ plane of the CMSSM scenario
           with $A=0$, $\tan\beta=10,50$ and neutralino LSP. Coloured
           regions represent the theoretical predictions for the stau
           flux at IceCube in the case of WB neutrino flux, 
           whereas solid vertical lines correspond to the results when
           then MPR flux is used. Regions where WMAP relic density is
           reproduced are represented by gridded areas.} 
  \label{fluxdeltam}
\end{figure}

This is indeed a very restrictive requirement which only leaves a
very narrow allowed band in the parameter space. In fact, such a small
mass-difference between the neutralino LSP and the stau 
implies a too small relic density for the neutralino when its mass is
small. The presence of staus with a sufficiently large lifetime is
therefore only compatible with neutralino dark matter above a certain
mass scale.

In order to illustrate this, we have represented in
Fig.\,\ref{fluxdeltam} the predicted flux of staus at the IceCube
detector in the $(\Delta m, \mstau)$ plane for an example of the CMSSM
with $A=0$, $\mu>0$ and $\tan\beta=10,\,50$, and for both a WB and a MPR
neutrino flux. As commented above, the
flux vanishes for $\Delta m\gsim2$~GeV. For $\Delta m\lsim2$~GeV the
flux increases as the stau mass decreases and can be as large as $2$
event yr$^{-1}$ for the MPR flux when the stau mass is close to its 
experimental lower bound. 
The WB flux is represented by means of a colour code, whereas the 
regions corresponding to different values of stau flux assuming a MPR
neutrino flux are delimited by solid lines.
The regions compatible with neutralino dark matter are represented on
the same plane by means of gridded areas (for those points where the
WMAP relic density is reproduced).

Compatibility of neutralino dark
matter with observable staus is only possible for $\tan\beta=10$ and
for $\mstau\approx 300$~GeV, thus implying a stau flux between $0.1$ 
and $0.5$ events yr$^{-1}$, in the optimistic MPR case or between 
$0.01$ and $0.1$ event yr$^{-1}$ if we work with a WB neutrino flux.
For larger values of $\tan\beta$ the region where the neutralino relic
density is compatible with WMAP
results  is shifted towards larger values of the stau mass 
and therefore is associated to much lower values of the stau flux. 
Once more, in spite of the good background discrimination,
these small fluxes would require, at best,
several years of data from IceCube.

\subsection{General supergravity with non-universal soft terms}

In the former Sections we have performed our calculations within the
framework of the CMSSM. We will now briefly explore how the
theoretical predictions for the stau flux vary when a more general
supergravity scenario with non-universal soft parameters is
considered. More specifically, we will try to identify possible
non-universal schemes that lead to an increase of the theoretical
predictions for the stau flux.

As we discussed above, the production rate of staus is very sensitive
to the value of the squark masses, the resulting flux increasing in
the presence of light squarks
\cite{Albuquerque2003,Albuquerque2006_1}.  
Likewise, a decrease in the slepton masses is also welcome in
order to increase their production cross-section.
Moreover,
as we can see in the Appendix, 
the expressions (\ref{sigma_charg}) corresponding
to the chargino-mediated sfermion production are proportional to 
the factor $|Z_+^{1k}\ Z_-^{1k}|^2$. This implies that
these contributions are enhanced when the lighter chargino is a pure 
wino state (i.e., $|Z_+^{1k}|\sim 1$).

A decrease of the low energy values for the 
squark masses, relative to the (right-handed) stau mass can be
obtained if the ratio of the gluino and bino mass parameters,
$M_3/M_1$, decreases at the GUT scale. 
The decrease in the squark masses induces a reduction of the $\mu$
parameter through the increase of the positive contributions to the
RGE of $m_{H_U}^2$. This in turn implies an unwanted enhancement of
the mixing in the
chargino mass matrix.  
Thus, in order to compensate
for the decrease in the $\mu$ parameter and have a pure wino as the
lighter chargino, as well as reducing its mass, 
we must also decrease the $M_2$ parameter at the GUT
scale. 
Smaller values of $M_2$ also imply a decrease in the masses for
left-handed sleptons. Although, as mentioned above, 
this is also potentially good to
increase their production cross-section, one
should bear in mind that eventually the sneutrinos can become almost
degenerate with the lighter stau and be long-lived. In that case
left-handed squarks would generally cascade down to the lightest
neutralino and this, in turn, mostly to sneutrinos which could
propagate through the Earth without decaying into staus, thereby
significantly attenuating the stau flux. We have avoided this
situation by making sure sneutrinos decay promptly (with a lifetime
smaller than $10^{-9}$ s),
thus constraining the non-universality in $M_2$.

Finally, we can also attempt to increase the $\mu$ parameter, thus
further enhancing the wino composition of the lightest chargino. This
can be done by introducing a departure from universality in the Higgs
mass parameters at the GUT scale. More specifically, we have
considered an 
increase of the value of $m_{H_U}^2$ with respect to the mass of the
rest of the scalars.

\begin{figure}[t]
  \begin{center}
    \epsfig{file=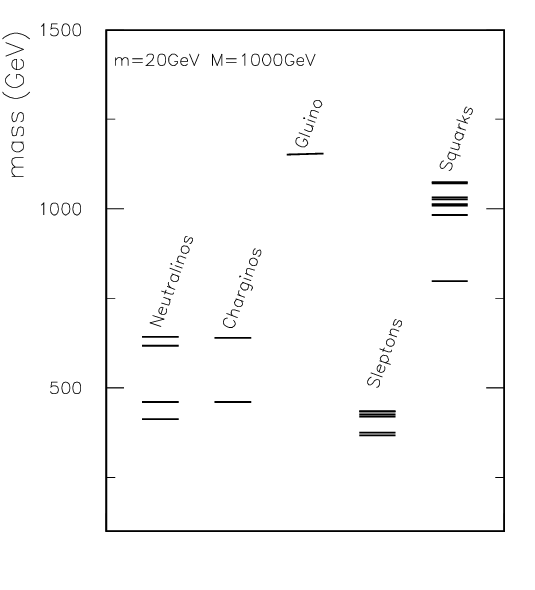,width=8.6cm}
  \end{center}
  \vspace*{-1cm}
  \captions{Supersymmetric spectrum corresponding to non-universal
    soft mass parameters as in Eq.(\ref{masses_nuniv}) with
    $M=1000$~GeV, $m=20$~GeV, $A=0$ and 
    $\tan\beta=10$.  
  }
  \label{spectrum_nuniv}
\end{figure}

As an specific example, following the above arguments we have taken
the following set of non-universalities in the gaugino and scalar
masses,
\begin{eqnarray}
  M_{2}&=&0.6\, M_1\,\nonumber\\
  M_{3}&=&0.5\, M_1\,\nonumber\\
  m_{H_D,\,Q,\,U,\,D,\,L,\,E}^2&=&m^2\,,\nonumber\\
  m_{H_U}^2&=&0.5\,m^2\,.
  \label{masses_nuniv}
\end{eqnarray}
The resulting supersymmetric spectrum is depicted in
Fig.\,\ref{spectrum_nuniv} for $M=1000$~GeV and $m=20$~GeV, clearly
evidencing the decrease in the masses of the squarks and heavy
sleptons relative to the
stau mass with respect to the universal case discussed in the previous
Sections and shown in Fig.\,\ref{spectrum_univ}.

Given the change in the SUSY spectrum, we have redone the analysis of
the decay chains that produce staus from squark and sleptons. We have
checked that in this example the relation between the resulting stau
energy and the parent squark and slepton reads 
\begin{eqnarray}
  \langle\estau\rangle\approx 0.8\,\langle\esl\rangle\ ;
  \quad \quad  
  \langle\estau\rangle\approx0.5\,\langle\esq\rangle\ .
\end{eqnarray}
The increase in both quantities with respect to the universal case
(see Eq.(\ref{energy-universal})) is due to the smaller mass
differences among the supersymmetric particles which cascade to the
lighter stau, which 
leads to a smaller suppression according to Eq.(\ref{e_stau}).
These relations approximately hold throughout the whole region of the
parameter space with stau LOSP.

\begin{figure}[!t]
  \begin{center}
    \epsfig{file=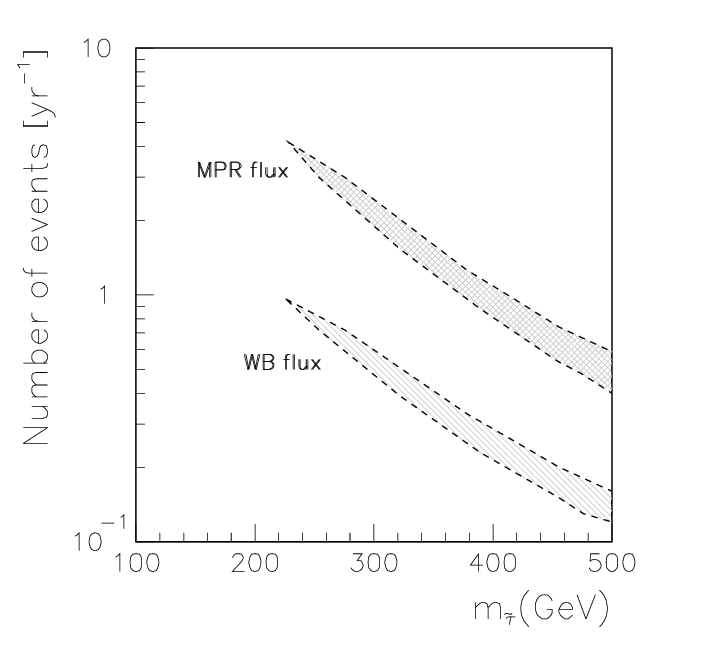,width=8.6cm}
  \end{center}
  \vspace*{-1cm}
  \captions{The same as in Fig.\ref{fluxewimp1} but for the example
    with non-universal soft mass parameters as in
    Eq.(\ref{masses_nuniv}). 
  }
  \label{fluxewimp-nuniv}
\end{figure}

The resulting theoretical predictions for the stau flux are
represented in
Fig.\,\ref{fluxewimp-nuniv}
as a function of the stau masses for both the WB and MPR flux for
$A=0$ and $\tan\beta=10$. A moderate increase is
observed with respect to the universal case of
Fig.\,\ref{fluxewimp1}. For example, the results using the WB flux
can reach now values of almost 1 event yr$^{-1}$, while those
corresponding to the MPR flux are approximately a factor three
larger.

\section{Conclusions}

We have explored the possibility of detecting long-lived staus at
neutrino telescopes, after their production inside the Earth through
the inelastic scattering of high energy neutrinos. More specifically,
we have calculated the production rate of a pair of staus in terms of
the high energy neutrino flux, 
evaluated their energy losses after traversing
the distance to the experiment, and taken into account the track
separation of the stau pair and the detector
efficiency in order to compute the theoretical predictions for the
resulting stau flux.
We have studied two generic scenarios in which long-lived staus are
naturally associated to a supersymmetric solution to the problem of
dark matter.

On the one hand, we have considered the case of superWIMP dark matter
in which the LSP is either the gravitino or the axino. Exploring the
areas of the CMSSM in which the stau is the lightest observable
particle we observed that the number of stau pairs 
is bounded to be below 1 event yr$^{-1}$ when the MPR neutrino flux is
used. These predictions decrease by approximately a factor four
for the WB flux. The largest values of the stau flux correspond
to small values of $\tan\beta$ and to 
regions with a small value of the common gaugino mass. These areas of
the parameter space can be consistent with viable gravitino dark
matter for light gravitinos ($m_{\widetilde G}\lsim 1$~GeV),
in order to avoid the stringent
BBN constraints, or axinos with mass 
$m_{\widetilde a}\sim{{\cal O}(1\,{\rm GeV})}$ and a small
reheating temperature, $T_R\sim200\,{\rm GeV}$.

On the other hand, we have explored the case with a neutralino LSP. 
In this scenario, the stau can be long-lived if the mass difference
with the neutralino is smaller than $\Delta
m\lsim2$~GeV.
Interestingly, the neutralino-stau coannihilation
region, where the WMAP relic density can be obtained, 
also occurs for small mass differences.
We have studied this possibility within the CMSSM, finding that once
more the largest results for the stau flux are obtained
for small values of the gaugino mass parameter, i.e., for light staus,
and increase when the rest of the spectrum (in particular 
the squark masses)
is also light. This favours low values of $\tan\beta$. For
$\tan\beta=10$, and using the MPR flux, 
an upper bound of approximately 2 events yr$^{-1}$ is
found for
staus with a mass of $m_{\stau}\sim 100$~GeV, which decreases by a factor
2 for $\tan\beta=50$. 
When the WMAP constraint is imposed on the neutralino relic abundance, 
compatibility with viable neutralino dark matter reduces the allowed
parameter space to a small range of stau masses. 
For example, with $\tan\beta=10$ one finds
$m_{\stau}\sim 300$~GeV, and the predicted stau flux is 
$0.1$ events yr$^{-1}$ .
When $\tan\beta$ increases the region compatible is shifted to heavier
staus and the resulting flux decreases significantly. 
Overall we have observed that 
the stau flux obtained in these supergravity scenarios is generally
smaller than those of low energy supersymmetric analyses.
This implies that, although stau events are distinguishable from the
dimuon backround, in the most optimistic scenarios
several years of data from IceCube are necessary to claim a positive
signal.

Finally, we have investigated the case of a general supergravity
theory in which the structure of the soft parameters is 
non-universal. We have observed that certain choices of
non-universalities which lead to wino-like charginos and 
a decrease of the squark and slepton masses
can account for a moderate increase of the resulting stau flux. 
More specifically, through 
a decrease of both the gluino and wino mass parameters with respect to
the bino mass and a decrease of the soft mass for the $H_u$ Higgs at
the GUT scale we have shown that the stau flux can increase by
approximately a factor three with respect to the results in the CMSSM.

\noindent{\bf Acknowledgements}

We thank K.Y. Choi and P. Slavich 
for very useful discussions. 
This work was
supported in part by the Spanish DGI of the
MEC under Proyecto Nacional FPA2006-01105, 
by the Comunidad de Madrid under Proyecto HEPHACOS, Ayudas de I+D
S-0505/ESP-0346, by the EU RTN
UniverseNet 
MRTN-CT-2006-035863, and by the ENTApP Network of the ILIAS project 
RII3-CT-2004-506222. 
The work of 
B. Ca\~nadas  has been partially supported by an INFN-Roma Tor
Vergata FAI grant for foreign researchers and thanks the INFN-Roma Tor
Vergata for their kind hospitality during the last stages of this work.
D.G. Cerde\~no is supported by the program ``Juan de la Cierva'' of
the Ministerio de Educaci\'on y Ciencia of Spain.
The work of C. Mu\~noz was supported
in part by the Spanish DGI of the
MEC under Proyecto Nacional FPA2006-05423,
and by the EU under the RTN program
MRTN-CT-2004-503369.

\clearpage
\appendix
\section{Sfermion production from neutrino inelastic scattering}
\label{appendix}

\begin{figure}[t]
  \begin{center}
    \begin{tabular}[h]{ccc}
      \includegraphics[width=.25\linewidth]{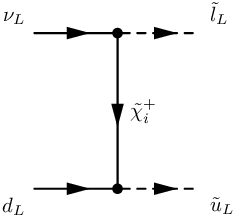} 
      &\quad\quad\quad\quad &
      \includegraphics[width=.25\linewidth]{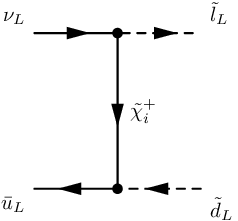} \\
      \rule{0pt}{25ex}
      \includegraphics[width=.25\linewidth]{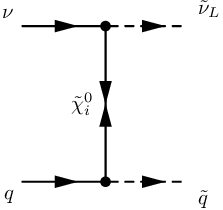} 
      &\quad\quad\quad\quad &
      \includegraphics[width=.25\linewidth]{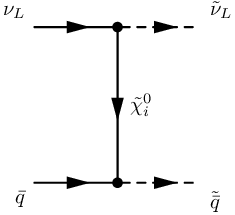} \\
    \end{tabular}
  \end{center}
  \captions{Feynman diagrams describing the different neutrino
    inelastic interactions. }
  \label{nud_chargino}
\end{figure}

The contribution from chargino exchange along a $t$-channel comprises
the diagrams shown in the upper row in Fig.~\ref{nud_chargino}. They 
lead to the following cross-sections,
\begin{align}
  \frac{d\sigma}{dt}\Big|_{\nu
    d}=&\frac{\pi\alpha^2}{2s_W^4}\frac{1}{s} 
  \left[\frac{m_{\widetilde\chi^+_k}|Z_U^{Ii}||Z_L^{Jj}||Z_+^{1k}||Z_-^{1k}|}
    {t-m_{\widetilde\chi^+_k}^2}\right]^2,\nonumber\\
  \frac{d\sigma}{dt}\Big|_{\nu
    \bar{u}}=&\frac{\pi\alpha^2}{2s_W^4}\frac{1}
       {s^2}\left[\frac{Z_-^{1k}Z_-^{1k}}
         {t-m_{\widetilde\chi^+_k}^2}\right]^2|Z_D^{Ii}||Z_L^{Jj}| 
       \left(tu-m_{\widetilde{d}_i}^2m_{\widetilde{l}_j}^2\right)\ ,
       \label{sigma_charg}
\end{align}
with notation for $Z$ matrices in 
agreement with that of \cite{Rosiek}, where $Z_+$ and $Z_-$ are
$2\times 2$ chargino mixing matrices, with $k=1,2$, and 
$Z_U$, $Z_D$ and $Z_L$ are $6\times6$ squark mixing 
matrices. Capital indices $I$
and $J$ are family indices running 
from 1 to 3, whereas indices $i$ and $j$ stand for mass eigenstates and run
from 1 to 6. Summation over all indices is implied.

For the diagrams with neutralino exchange along a $t$-channel, 
shown on the lower row of Fig.~\ref{nud_chargino}, we make the
following definitions, 
\begin{displaymath}
  \begin{array}{ll}
    {N}_i^{\nu_L}= c_WZ_N^{2i}-s_WZ_N^{1i}   \,,         &
    \nonumber \cr 
    {N}_i^{u_L}  = c_WZ_N^{2i}+\frac{1}{3}s_WZ_N^{1i}  \,,  \qquad& 
    {N}_i^{u_R}=\frac{4}{3}s_WZ_N^{1i}  \,,  \nonumber \cr
    {N}_i^{d_L}  =-c_WZ_N^{2i}+\frac{1}{3}s_WZ_N^{1i} \,,   \qquad&
    {N}_i^{d_R}=-\,\frac{2}{3}s_WZ_N^{1i} \,.  \nonumber 
  \end{array}
\end{displaymath}
The total cross-section for interaction with quarks, expressed in terms of 
mass eigenstates, reads
\begin{align}
  \frac{d\sigma}{dt}\Big|_{\nu q}=
  \frac{\pi\alpha^2}{8s_W^4c_W^4}\frac{1}{s^2}
  \left\{s\left[\frac{m_{\widetilde\chi^0_k}{N}_k^{q_L}
      {N}_i^{\nu_L}|Z_{U/D}^{Ii}|}{(t-m_{\widetilde\chi^0_k}^2)}\right]^2+
  \sum_i\left[\frac{{N}_i^{q_R}{N}_i^{\nu_L}
      |Z_{U/D}^{(I+3)i}|}{(t-m_{\widetilde\chi^0_i}^2)}\right]^2
  (tu-m_{\widetilde{q_i}}^2m_{\widetilde{\nu}}^2)\right\}
  \,,\nonumber
  \\ 
  \frac{d\sigma}{dt}\Big|_{\nu
    \bar{q}}=\frac{\pi\alpha^2}{8s_W^4c_w^4}\frac{1}{s^2}
  \left\{s\left[\frac{m_{\widetilde\chi_i^0}^2{N}_i^{u_R/d_R}
      {N}_i^{\nu_L}|Z_{U/D}^{(I+3)i}|}{(t-m_{\widetilde\chi^0_i}^2)}
    \right]^2+\left[\frac{{N}_i^{u_L/d_L}
      {N}_i^{\nu_L}|Z_{U/D}^{Ii}|}{(t-m_{\widetilde\chi^0_i}^2)}\right]^2
  (tu-m_{\widetilde{q}^2\widetilde{l}^2})\right\} 
\end{align} 
These parton-level cross-sections are then convoluted with their 
corresponding Parton Distribution Functions, which we extract from the
Cteq6PDF
package \cite{cteq6}.

To calculate the fraction of the parent's energy carried by the stau,
we use the general formula 
\begin{equation}
  E_{\widetilde{\tau}}^{LAB}=\frac{E_{parent}^{LAB}}{2^n}\prod_{i=1,n}
  \left(1+\frac{m_{i}^2}{m_{i-1}^2}\right)\ ,
  \label{e_stau}
\end{equation}
where $n$ is the number of intermediate states, $i=0$ corresponds to
the parent squark or slepton, and $i=n$ corresponds to the stau.

\begingroup\raggedright\endgroup

\end{document}